\documentclass{iaus}
\usepackage{graphicx}

\title[LSST and NEOs] 
{LSST: Comprehensive NEO Detection, Characterization, and Orbits}

\author[Ivezi\'{c} et al.]   
{\v{Z}eljko Ivezi\'{c}$^1$, J. Anthony Tyson$^2$, Mario Juri\'{c}$^3$, Jeremy
  Kubica$^4$, Andrew Connolly$^5$, Francesco Pierfederici$^6$, 
  Alan W. Harris$^7$, Edward Bowell$^8$,
  and the LSST Collaboration$^9$
  \thanks{}
}
\affiliation{
$^1$Department of Astronomy, University of Washington, Seattle, WA 98155, USA 
    \break email:  ivezic@astro.washington.edu \\[\affilskip]
$^2$Department of Physics, University of California, Davis, CA 95616, USA
$^3$Princeton University Observatory, Princeton, NJ 08544, USA
$^4$Google Inc., 1600 Amphitheatre Parkway, Mountain View, CA 94043, USA 
$^5$Department of Astronomy, University of Washington, Seattle, WA 98155, USA
$^6$LSST Corporation, 4703 E. Camp Lowell Drive, Suite 253, Tucson, AZ 85712, USA
$^7$Space Science Institute, 4603 Orange Knoll Ave., La Canada, CA 91011-3364, USA
$^8$Lowell Observatory, 1400 W. Mars Hill Rd., Flagstaff, AZ 86001, USA
$^9$ www.lsst.org
}

\pubyear{2006}
\volume{236}  
\pagerange{1--10}
\date{Received Sep 30, 2006; and in revised form Nov 17, 2006}
\jname{IAU Symposium 236 ``Near Earth Objects, Our Celestial Neighbors: Opportunity And Risk"}
\editors{Andrea Milani, Giovanni B. Valsecchi \& David Vokrouhlicky, eds.}
\begin{document}

\newcommand\x         {\hbox{$\times$}}
\newcommand\othername {\hbox{$\dots$}}
\def\eq#1{\begin{equation} #1 \end{equation}}
\def\eqarray#1{\begin{eqnarray} #1 \end{eqnarray}}
\def\eqarraylet#1{\begin{mathletters}\begin{eqnarray} #1 %
                  \end{eqnarray}\end{mathletters}}
\def\mic              {\hbox{$\mu{\rm m}$}}
\def\about            {\hbox{$\sim$}}
\def\Mo               {\hbox{$M_{\odot}$}}
\def\Lo               {\hbox{$L_{\odot}$}}
\def\sun              {\hbox{$\odot$}}
\def\comm#1           {{\tt (COMMENT: #1)}}
\def\sm#1           {{\tt (MACRO: #1)}}
\def\deg              {$^\circ$}
\def\mone		{{$^{-1}$}}
\def\masyr             {{~mas~yr$^{-1}$}}
\def\kms		{{~km~s$^{-1}$}}
\def\etal{{\it et al.}\xspace}
\def\eg{{\it e.g.}\xspace}
\def\ie{{\it i.e.}\xspace}

\def\Figure#1#2[#3] {
\centerline{
\scalebox{#3}{
\includegraphics{#1.#2}
}}}

\maketitle

\begin{abstract}
The Large Synoptic Survey Telescope (LSST) is currently by far the 
most ambitious proposed ground-based optical survey. With initial 
funding from the National Science Foundation (NSF), Department of 
Energy (DOE) laboratories, and private sponsors, the design and 
development efforts are well underway at many institutions, including 
top universities and leading national laboratories. Solar System mapping 
is one of the four key scientific design drivers, with emphasis on 
efficient Near-Earth Object (NEO) and Potentially Hazardous Asteroid 
(PHA) detection, orbit determination, and characterization.

The LSST system will be sited at Cerro Pachon in northern Chile. In 
a continuous observing campaign of pairs of 15 second exposures of its 
3,200 megapixel camera, LSST will cover the entire available sky every 
three nights in two photometric bands to a depth of V=25 per visit (two 
exposures), with exquisitely accurate astrometry and photometry. Over the 
proposed survey lifetime of 10 years, each sky location would be visited 
about 1000 times, with the total exposure time of 8 hours distributed over 
several broad photometric bandpasses. The baseline design satisfies strong 
constraints on the cadence of observations mandated by PHAs such as closely 
spaced pairs of observations to link different detections and short exposures 
to avoid trailing losses. Equally important, due to frequent repeat visits 
LSST will effectively provide its own follow-up to derive orbits for detected 
moving objects. 

Detailed modeling of LSST operations, incorporating real historical weather 
and seeing data from Cerro Pachon, 
shows that LSST using its baseline design cadence could find 90\% of the PHAs 
with diameters larger than 250 m, and 75\% of those greater than 140 m within 
ten years. However, by optimizing sky coverage, the ongoing simulations suggest 
that the LSST system, with its first light in 2013, can reach the 
Congressional mandate of cataloging 90\% of PHAs larger than 140m by 2020. 
In addition to detecting, tracking, and determining orbits for these PHAs, 
LSST will also provide valuable data on their physical and chemical 
characteristics (accurate color and variability measurements), constraining 
PHA properties relevant for risk mitigation strategies.

In order to fulfill the Congressional mandate, a survey with an etendue of
at least several hundred m$^2$deg$^2$, and a sophisticated and robust data 
processing system is required. It is fortunate that the same hardware, 
software and cadence requirements are driven by science unrelated to NEOs: 
LSST reaches the threshold where different science drivers and different 
agencies (NSF, DOE and NASA) can work together to efficiently achieve seemingly 
disjoint, but deeply connected, goals. 
\end{abstract}

\section{      {\bf Introduction   }     }

\subsection{         The Challenges     }

We are immersed in a swarm of Near Earth Asteroids (NEAs) whose orbits approach that of 
Earth. About 20\% these, the potentially hazardous asteroids (PHAs), are in orbits that 
pass close enough to Earth's orbit ($<$0.05 AU) that perturbations with time scales of 
a century can lead to intersections and the possibility of collision. 
Beginning in 1998, NASA set as a goal the discovery 
within 10 years of 90\% of the estimated 1000 NEAs with diameters greater than 1 km. 
It is expected that ongoing surveys will in fact discover about 80\% of these large 
($>$1 km) NEAs by 2008 (Jedicke et al. 2003). However, this mission has been
recently extended by the US Congress. The following text became law as part of
the NASA Authorization Act of 2005 
passed by the Congress on December 22, 2005, and subsequently signed by the President:

``The U.S. Congress has declared that the general welfare and security of the United States
require that the unique competence of NASA be directed to detecting, tracking, cataloguing,
and characterizing near-Earth asteroids and comets in order to provide warning and
mitigation of the potential hazard of such near-Earth objects to the Earth. 
The NASA Administrator shall plan, develop, and implement a Near-Earth Object Survey
program to detect, track, catalogue, and characterize the physical characteristics of near-
Earth objects equal to or greater than 140 meters in diameter in order to assess the threat of
such near-Earth objects to the Earth. It shall be the goal of the Survey program to achieve
90\% completion of its near-Earth object catalogue (based on statistically predicted
populations of near-Earth objects) within 15 years after the date of enactment of this Act.''

Ground-based optical surveys are the most efficient tool for comprehensive 
NEO detection, determination of their orbits and subsequent tracking. A
survey capable of extending these tasks to NEOs with diameters as small as 
140 m requires a large telescope, a large field of view (FOV) and a sophisticated
data acquisition, processing and dissemination system.

A 140-meter object with a typical albedo (0.1), positioned in the main asteroid 
belt (at a heliocentric distance of 2.5 AU), and observed at opposition will 
have an apparent visual Johnson magnitude of V$\sim$25. In order to detect 
such a faint object with at least 5$\sigma$ significance, in an exposure not 
longer than 30 seconds to prevent trailing losses (the maximum exposure
time is even shorter for NEOs observed at low elongations around so-called
``sweet spots''), a 10-meter class telescope 
is needed even at the best observing sites. A large FOV, of the 
order 10 square degrees, and short slew time are required to enable repeated 
observations of a significant sky fraction with a sufficient frequency. With 
such observing cadence and depth, observations will necessarily produce tens 
of terabytes of imaging data per night. In order to recognize NEOs, determine 
their orbits and disseminate the results to the interested communities in a
timely manner, a powerful and fully automated data system is mandatory. 

These considerations strongly suggest that in order to fulfill the Congressional 
mandate, a system with a high etendue (or throughput, defined as the product 
of the aperture area and field-of-view area), of at least several 
hundred m$^2$deg$^2$, is required. The LSST system has nearly two orders of 
magnitude larger etendue than that of any existing facility (Tyson 2002), 
and is the only facility that can detect 140-meter objects in the main
asteroid belt in less than a minute.

\begin{figure}[!t]
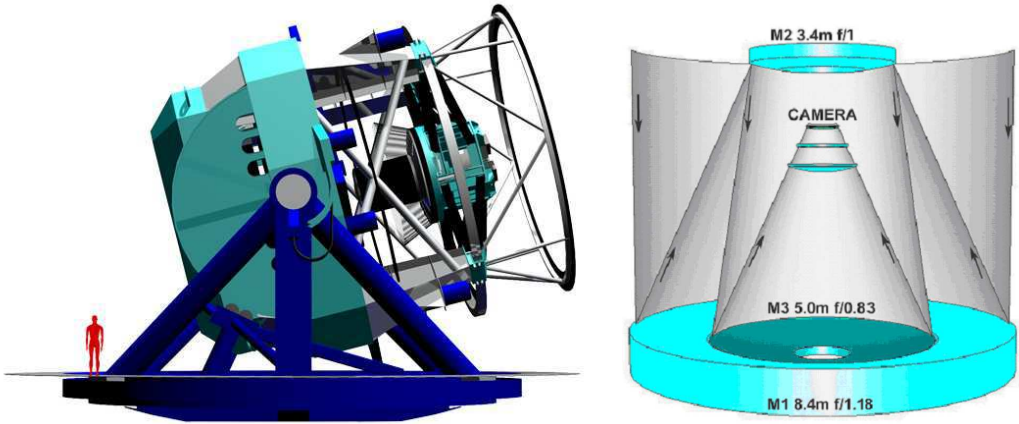

\phantom{x}
\vskip -1.7in
\hskip -1.1in 
\Figure{telescope2S}{ps}[0.55]
\vskip -4.7in
\hskip 1.6in
\Figure{opticsS}{ps}[0.3]
\vskip -0.7in
\caption{The left panel shows baseline design for LSST telescope, current as of 
April 2006. The telescope will have an 8.4-meter primary mirror, and a 
10-square-degree field of view. The right panel shows LSST baseline optical design 
with its unique monolithic mirror: the primary and tertiary mirrors are coplanar 
and their surfaces will be polished into single substrate.}
\label{tel}
\end{figure}

\begin{figure}[!h]
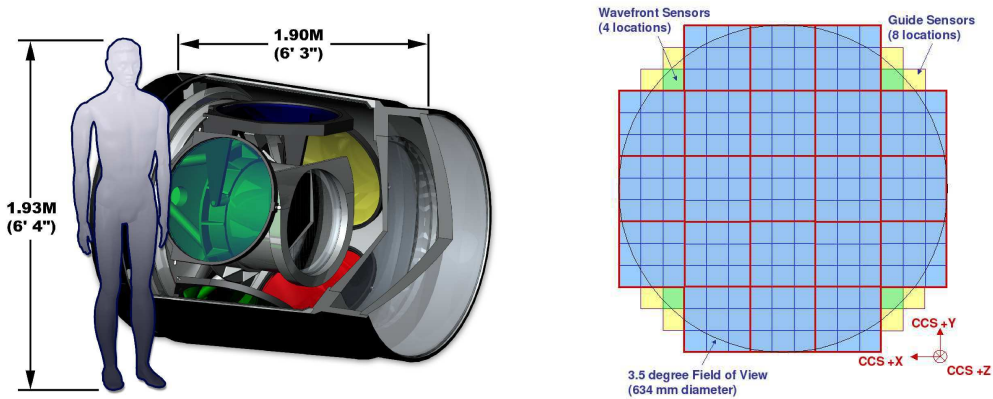

\vskip -0.7in
\phantom{x}
\hskip -1.3in 
\Figure{camera1S}{ps}[0.39]
\vskip -3.9in
\hskip 1.5in
\Figure{fovS}{ps}[0.3]
\vskip -0.7in
\caption{
The left panel shows LSST camera with person to indicate scale size. The
camera is positioned in the middle of the telescope and will include a filter 
mechanism and shuttering capability.
The right panel shows LSST focal plane. Each cyan square represents one
4096x4096 pixel large sensors. Nine sensors are assembled together in a raft. 
There are 189 science sensors, each with 16.8 Mpix, for the total pixel count 
of 3.2 Gpix.}
\label{fov}
\end{figure}

\subsection{            The LSST Drivers    }

Three recent committees comissioned by the National Academy of 
Sciences\footnote{
  Astronomy and Astrophysics in the New Millennium, NAS 2001; 
  Connecting Quarks with the Cosmos: Eleven Science Questions for the New Century, NAS 2003; 
  New Frontiers in the Solar System: An Integrated Exploration Strategy, NAS 2003.   
}
concluded that a dedicated wide-field imaging telescope with an effective aperture 
of 6--8 meters is a high priority for US planetary science, astronomy, and physics 
over the next decade. The LSST system described here will be a large, 
wide-field ground based telescope designed to obtain sequential images covering the 
entire visible sky every few nights. The current baseline design allows us to do so 
in two photometric bands every three nights. 
  
The survey will yield contiguous overlapping imaging of $\sim$20,000 square degrees 
of sky in at least five optical bands covering the wavelength range 320--1050 nm.
Detailed simulations that include measured weather statistics and a variety 
of other effects which affect observations predict that each sky location can be 
visited about 100 times per year, with two 15 sec exposures per visit.  

The range of scientific investigations which would be enabled by such a 
dramatic improvement in survey capability is extremely broad. The main
science themes that drive the LSST system design are
\begin{enumerate}
\item Constraining Dark Energy and Matter
\item Taking an Inventory of the Solar System
\item Exploring the Transient Optical Sky
\item Mapping the Milky Way
\end{enumerate}

In particular, the detection, characterization and orbital determination
of Solar System objects impact the requirements on relative and absolute
astrometric accuracy (10 milliarcsec and 50 milliarcsec, respectively,
for sources not limited by photon statistics), and drive the detailed
cadence design, discussed further below.

\subsection{     The LSST  Reference Design    }

The LSST reference design\footnote{More details about LSST system are available at 
http://www.lsst.org.}, with an 8.4 m diameter primary mirror, standard
filters ($ugrizY$, 320 -- 1050 nm), and current detector performance, reaches 
24th V mag in 10 seconds\footnote{An LSST exposure time calculator has been 
developed and is publicly available at 
http://tau.physics.ucdavis.edu/etc/servlets/LsstEtc.html.}. 
With an effective aperture of 6.5m and 9.6 square degree field of view, LSST 
has an etendue of 320 m$^2$deg$^2$. This large
etendue is achieved in a novel three-mirror design (modified Paul-Baker) with 
a very fast f/1.25 beam, and a 3.2 gigapixel camera (with 0.2 arcsec large
pixels). 
The baseline designs for telescope and camera are shown in Figs.~\ref{tel} 
and \ref{fov}. The LSST telescope will be sited at Cerro Pachon, Chile.

\section{         \bf The LSST PHA Survey  }

\subsection{          The PHA Survey Requirements    }

The search for PHAs puts strong constraints on the cadence of observations, 
requiring closely spaced pairs of observations two or preferably three times 
per lunation in order to link observations unambiguously and derive orbits. 
Individual exposures should be shorter than about 20 sec each to minimize 
the effects of trailing for the majority of moving objects. Because of the 
faintness and the large number of PHAs and other asteroids that will be 
detected, LSST must provide the follow-up required to derive orbits rather 
than relying, as current surveys do, on separate telescopes. The observations 
should be preferentially obtained within $\pm15$ degrees of the Ecliptic,
with additional all-sky observations to increase the completeness at the 
small size limit (because the smaller asteroids must be closer to the Earth 
in order to be visible, and therefore are more nearly isotropically distributed).

The images should be well sampled to enable accurate astrometry, with absolute 
accuracy not worse than 0.1 arcsec. There are no special requirements on filters, 
although bands such as V and R that offer the greatest sensitivity are preferable.  
The images should reach a depth of at least 24 (5$\sigma$ for point sources) 
in order to probe the $\sim140$ m size range at main-belt distances. 
Based on recent photometric measurements of asteroids by the Sloan Digital Sky 
Survey, the photometry should be better than 1-2\% to allow for color-based 
taxonomic classification and light-curve measurements.

\subsection{          The PHA Survey Baseline Design   }

The baseline design cadence is based on two revisits closely separated in time 
(15-60 min) to enable a robust and simple method for linking main-belt asteroids 
(MBAs). Their sky surface density is about two orders of magnitude higher than 
the expected density of PHAs, and thus MBAs must be efficiently and robustly 
recognized in order to find PHAs. MBAs move about 3-18 arcmin in 24 hours, which 
is larger than their typical nearest neighbor angular separation at the depths probed by 
LSST (2.3 arcmin on the Ecliptic). Two visits closely separated in time (``re-visits") enable 
linking based on a simple search for the nearest moving neighbors, with a false 
matching rate of only a few percent. 

The present planned observing strategy is to ``visit" each field (9.6 sq. deg.) 
with two back-to-back exposures of $\sim$15 sec, reaching to at least $V$ magnitude 
of $\sim$24.8. Two such visits will be spaced in time by about half an hour.
Each position in the sky  will be visited several times during a month, spaced 
by a few days. This cadence will result in orbital parameters for several million 
MBAs, with light curves and color measurements for a substantial fraction of each population. 
Compared to the current data available, this would represent a factor of 10 to 100 
increase in the numbers of orbits, colors, and variability of the two classes of 
object. The large LSST MBA sample will enable detailed studies of the dynamical 
and chemical history of the Solar System.

\begin{figure}[t]
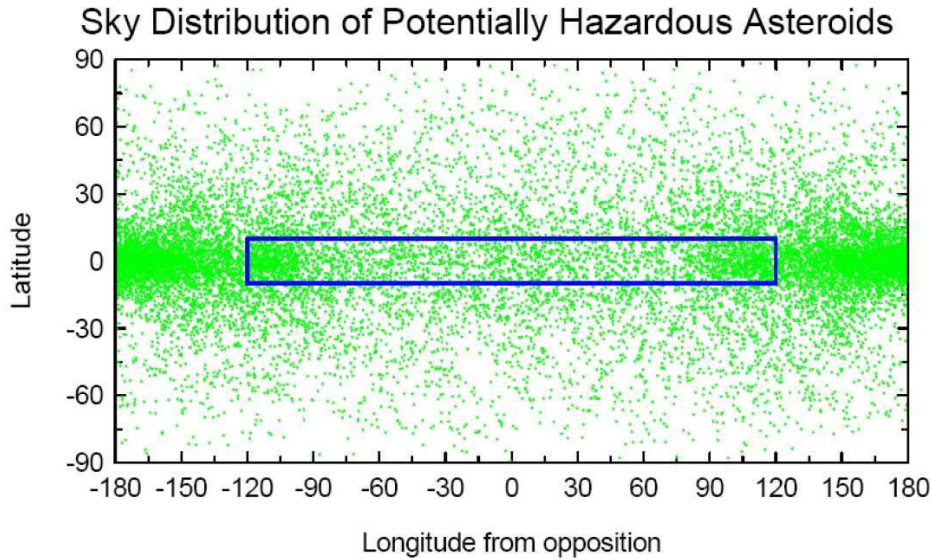

\vskip -2.0in
\Figure{phaS}{ps}[0.70]
\vskip -2.5in
\caption{Distribution of PHAs (dots) on the sky in ecliptic coordinate
system for objects with $V<H+2$ (e.g. $V<24$ for $H=22$ objects). 
The box extending $\pm$10$^\circ$ in latitude and $\pm$120$^\circ$ 
in longitude from the opposition point represents a region targeted by 
most current surveys because it yields a good efficiency in surveying for PHAs. 
However, the baseline LSST PHA survey will not be limited to this region, and 
will greatly benefit from its frequent all-sky coverage with faint magnitude 
limits.
}
\label{PHAsky}
\end{figure}

\subsection{                   The PHA Survey Simulations                  }

\begin{figure}[t]
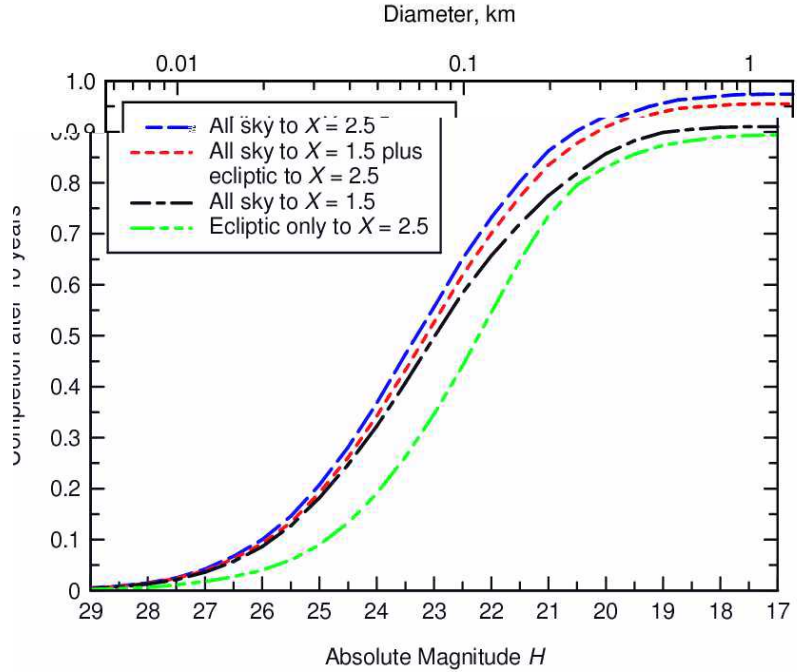

\vskip -1.4in
\Figure{complint2S}{ps}[0.6]
\vskip -1.8in
\caption{Completeness for PHAs for LSST baseline design cadence as function of 
absolute magnitude $H$ (or, equivalently, object's size, as marked on top). 
Different curves correspond to different sky tiling strategies ($X$ is
airmass). LSST baseline design 
cadence results in 90\% completeness for PHAs with diameters larger than 250 m, and 
75\% completeness for those greater than 140 m, after ten years. Ongoing simulations
suggest that, with additional optimization of the observing cadence, LSST can achieve 
90\% completeness for PHAs with diameters larger than 140 m.}
\label{PHAcompleteness}
\end{figure}

\begin{figure}[t]
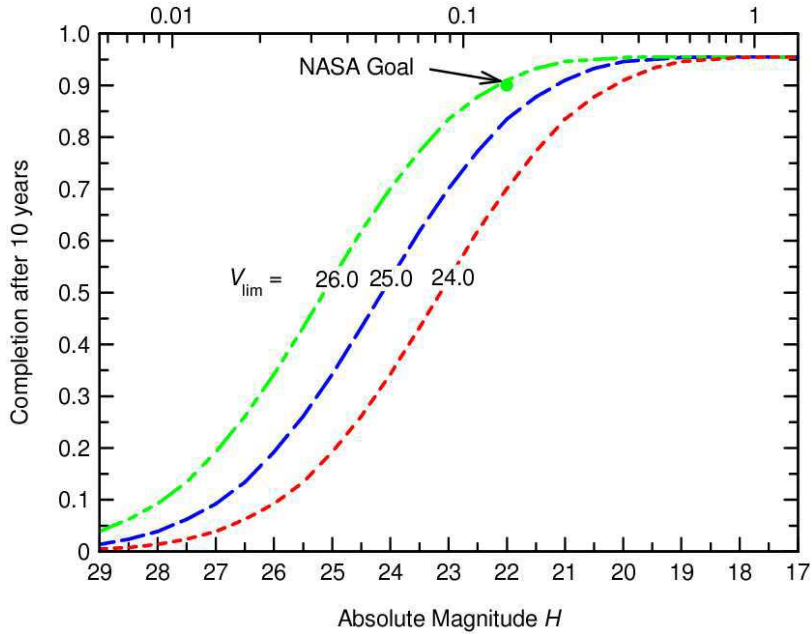

\vskip -1.4in
\Figure{compl25cS}{ps}[0.6]
\vskip -1.8in
\caption{10-year completeness for PHAs for various LSST single visit limiting magnitudes as function of 
absolute magnitude $H$ (or, equivalently, object's size, as marked on top). 
Different curves correspond to different limiting magnitudes per visit. LSST baseline design 
cadence corresponds closely to the middle curve.  If all the data in other bands were used in the
search for PHAs, then the completeness rises by 10\% reaching the goal of 90\% completeness for PHAs 
with diameters larger than 140 m. For PHAs with large collision cross-section this actually corresponds to
nearly 95\% completeness, as described in the text.}
\label{compl25}
\end{figure}

The performance of the baseline design cadence is studied and quantified using 
detailed simulations (developed by A. Harris and E. Bowell), including real 
historical weather and seeing data from Cerro Pachon (LSST site). The survey 
is simulated by generating a set of 1000 synthetic orbital elements that match the 
distribution of discovered objects in the large size range where present surveys are 
essentially complete. Positions and magnitudes in the sky are computed at five day 
intervals for ten years. An example of the instantaneous distribution of the 
simulated sample on the sky is shown in Figure~\ref{PHAsky}. 

The resulting sample is then ``filtered" according to assumed sky coverage and cadence 
pattern, limiting magnitude of survey instrument, visibility constraints, and magnitude
loss according to observing conditions. It is assumed that detections on two days out 
of three days observed in a month suffice for orbit determination (justification for
this assumption is discussed further below). This selection is repeated as a function of 
absolute magnitude, $H$, each time tabulating how many of the 1000 objects are ``discovered". 
The completeness as a function of absolute magnitude (i.e. size) for the baseline design 
cadence is shown in Figure~\ref{PHAcompleteness}. 

These simulations suggest that it is essential to cover the ecliptic band to as great 
an elongation as possible.  We find that a band extending $\pm$10$^\circ$ in latitude and 
$\pm$120$^\circ$  in longitude from the opposition point (see the blue box in \
Figure~\ref{PHAsky}) yields fairly good efficiency in surveying for PHAs.  This is true 
because even objects in highly inclined orbits must pass through the ecliptic sometime.  
Nevertheless, adding coverage of the rest of the sky improves the survey 
efficiency at the small size limit markedly.

The LSST baseline design cadence can achieve, over 10 years, a completeness of 90\% 
for objects larger than $\sim$250 m diameter, and 75\% completeness for those greater 
than 140 m. However, ongoing simulations show that by using all available data, 
optimizing filter choice and 
operations, LSST would be capable of just reaching a completeness of 90\% for PHAs 
larger than 140 m in ten years. For example, adjusting the depth of each visit affects
completeness as shown in Figure~\ref{compl25}. Simply including all the relevant data
from other filter bands can also increase completeness. This could result in an increase of 10\%
for 140 m PHAs, moving up from the middle curve to the top curve.  Moreover, note that none
of the curves saturate at 100\%.  That remaining 5\% represents PHAs with mainly low collision
cross section -- thus they pose no substantial hazard.  In effect, the completeness for truly
hazardous PHAs of $>$140m size is already close to 90\% in the LSST baseline cadence.
Due to its large etendue, LSST system enables sufficiently frequent sky coverage to assure
multiple detections per lunation even at the 140 m limit.

Perhaps the most important insight from these simulations is the realization 
that a system with an etendue of at least several hundred m$^2$deg$^2$ is 
mandatory for fulfilling the Congressional mandate to catalog 90\% of PHAs
larger than 140m.

\subsection{    From Many Detections to Orbits    }

The LSST PHA survey simulations described above assume that detections on 
two days out of three days observed in a month suffices for orbit 
determination. The linkage of individual detections and orbit determination
will be a formidable task for any large area survey. The increase in data 
volume associated with LSST will make the extraction of tracks and orbits of 
asteroids  from the underlying clutter a significant computational challenge.

The combinatorics involved in linking multiple observations of $\sim$10$^6$
sources spread out over several nights will overwhelm naive linear and quadratic 
orbit prediction schemes. Tree-based algorithms for multihypothesis testing 
of asteroid tracks can help solve these challenges by providing the necessary 
1000-fold speed-ups over current approaches while recovering 99\% of the 
underlying objects. 

In addition, observations of asteroids are often incomplete. Sources fall 
below the detection threshold in one or more of the series of observations. 
Weather results in incomplete sampling of the temporal data. Noise introduces
photometric and astrometric uncertainties. Combined, these observational
constraints can severely limit our ability to survey large areas by
requiring that we revist a given pointing on the sky more often than
necessary. The LSST's approach is designed to be robust to these
effects.

LSST will use a three stage process to find new moving objects (Kubica 2005,
Kubica et al. 2005). In the first 
stage intra-night associations are proposed by searching for detections
forming linear ``tracklets.''  By using loose bounds on the linear fit and 
the maximum rate of motion, many erroneous initial associations can be ruled
out. In the second stage, inter-night associations are proposed by searching 
for sets of tracklets forming a quadratic trajectory.  Again, the algorithm 
can efficiently filter out many incorrect associations while retaining most 
of the true associations. However, the use of a quadratic approximation means 
that a significant number of spurious associations still remains. In the
third stage, initial orbit determination and differential corrections
algorithms are used to further filter out erroneous associations by rejecting 
associations that do not correspond for a valid orbit.  Each stage of this
strategy thus significantly reduces the number false candidate associations 
that the later and more expensive algorithms need to test. 

To implement this strategy, the LSST team has developed, in a collaboration with 
the Pan-STARRS project (Kaiser et al. 2002), a pipeline based on multiple 
kd-tree data structures (Barnard et al. 2006).
These data structures provide an efficient way of 
indexing and searching large temporal data sets. Implementing a variable tree 
search we can link sources that move between a pair of observations, merge these 
tracklets into tracks spread out over tens of nights, accurately predict where a 
source will be in subsequent observations and provide a set of candidate asteroids 
ordered by the likelihood that they have valid asteroid tracks. Tested on
simulated data, this pipeline recovers 99\% of correct tracks for NEOs and
MBAs, and requires less than a day of CPU time to analyze a night's worth of 
data. This represents a several thousand fold increase in speed over a naive 
linear search. It is noteworthy that comparable amounts of CPU time are spent 
on kd-tree based linking step (which is very hard to parallelize) and on 
posterior orbital calculations to weed out false linkages (which can be
trivially parallelized).  

Given the predicted astrometric accuracy of the survey (per exposure: 10 milliarcsec relative
and 50 milliarcsec absolute, root-mean-square scatter per coordinate for sources
not limited by photon statistics) and the efficiency of the tracking algorithms
described above, LSST can lose all but two observations in one 24 hour period and 
still recover sufficient information to determine the tracklet (linking pairs
of re-visits) required to initiate a search for an asteroid. Pairs of these
tracklets can be observed as 
infrequently as once every eight nights and still result in a candidate asteroid 
being identified with an accuracy of 90\% or greater.

\subsection{    PHA Characterization   }

\begin{figure}[t]
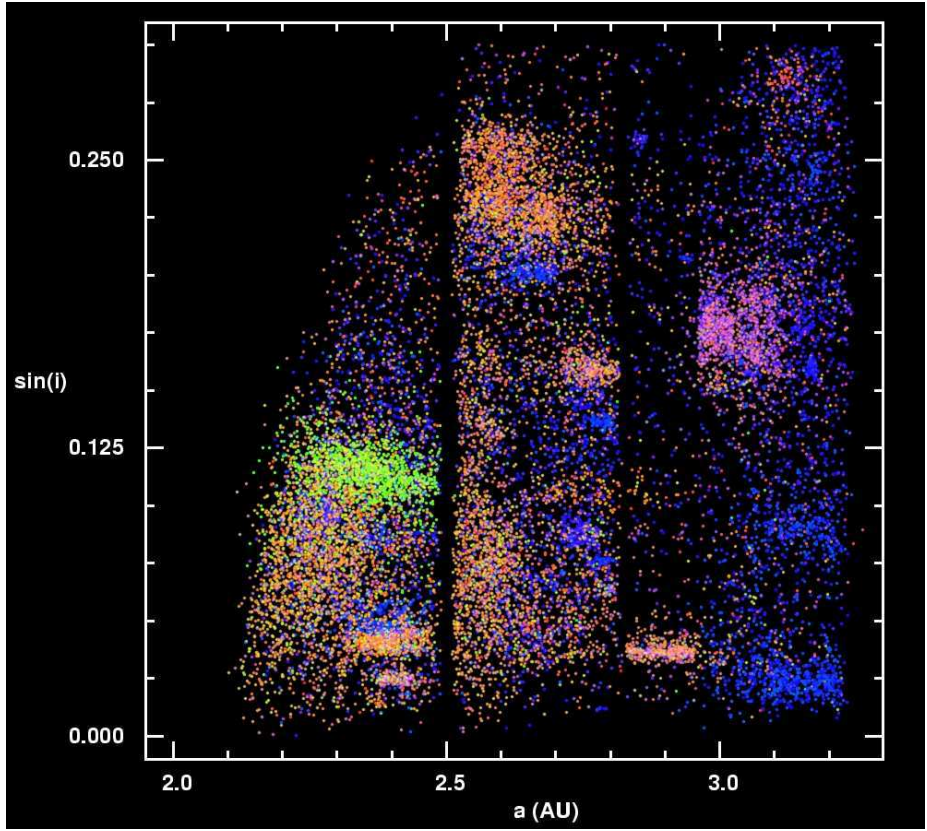

\vskip -1.4in
\Figure{a_sini_prop4cS}{ps}[0.69]
\vskip -1.8in
\caption{The distribution of $\sim$30,000 asteroids from the SDSS Moving 
Object Catalog (Available from http://www.sdss.org/science/index.html) in the plane spanned 
by the proper orbital inclination and semi-major axis (approximately, the x axis is 
proportional to the distance from the Sun, and y axis is proportional to the distance 
from the orbital plane) computed by Milani \& Kne\v{z}evi\'{c} (1992). The
dots are colored according to their {\it measured} SDSS 
colors. The clusters of points are Hirayama's dynamical families, proposed to represent 
remnants of larger bodies destroyed in collisions. SDSS data vividly demonstrate
a strong correlation between dynamics and colors, in support of Hirayama's hypothesis.
LSST will extend this map to a three times smaller size limit for more than ten
times larger sample, and will also provide structural information by measuring
light curves for most objects. LSST will make it possible to produce analogous
plot for about the same number of trans-Neptunian objects as shown here for
main-belt asteroids.}
\label{MBAsdss}
\end{figure}

LSST will not only obtain orbits for PHAs, but will also provide valuable data 
on their physical and chemical characteristics, constraining the PHA properties 
relevant for risk mitigation strategies. LSST will measure accurate colors for a substantial 
fraction of detected moving objects, thereby allowing studies of their surface 
chemistry, its evolution with time, and of dynamical (collisional) evolution.
As a recent example obtained by a modern large-area multi-color survey, 
Figure~\ref{MBAsdss} shows a correlation between orbital elements and optical
colors for MBAs measured by the Sloan Digital Sky Survey (Ivezi\'{c}
et al. 2001). The clusters of 
points visible in the figure are Hirayama's dynamical families, proposed to represent 
remnants of larger bodies destroyed in collisions. SDSS data vividly demonstrate
a strong correlation between dynamics and colors, in support of Hirayama's hypothesis.
LSST will extend this map to a three times smaller size limit, and will also provide 
structural information by measuring light curves for the majority of objects. The
variability information carries important information about the physical state of an 
asteroid (e.g. solid body vs. a rubble pile; Pravec \& Harris 2000), and these new data 
will constrain the size-strength relationship, which is a fundamental quantity that 
drives the collisional evolution of the asteroid belt.

Inevitably, some PHAs will be sufficiently interesting, based on provisional 
orbits from LSST data, that they should be continuously followed to ascertain their
detailed light-curve (rubble pile or solid body). 
Through Las Cumbres Observatory, one of the LSSTC member institutions, 
we will have access to a worldwide array of 2m-class telescopes for this continuous 
tracking. These telescopes will be instrumented with identical detectors and filters to
those on LSST in a dichroic camera capable of taking simultaneous images at
multiple wavelengths. This color ``movie" will aid characterization of these PHAs.

\section{                Conclusions            }

LSST is in a unique position to fulfill the Congressional mandate
to reach 90\% completion for 140 m large NEOs because such a survey
requires a large telescope to achieve necessary depths (at least 
$V\sim24.5$) with exposure times not longer than 15-20 seconds (to avoid 
trailing losses), a large field of view to be able to scan the sky at a 
required pace, and a sophisticated data acquisition, processing  and 
dissemination system to handle billions of detectable stars and galaxies. 

Detailed simulations suggest that 
the LSST baseline design cadence will achieve, over 10 years, a completeness 
of 90\% for objects larger than $\sim$250 m diameter, and 70\% completeness 
for those greater than 140 m. Ongoing simulations suggest that with further 
minor optimization of the baseline cadence LSST would be capable of reaching 
the Congressional target completeness of 90\% for PHAs larger than 140 m. 
In addition, LSST will not only obtain orbits for PHAs, but will also provide 
valuable data on their physical and chemical characteristics, constraining the 
PHA properties related to risk mitigation. 

In summary, LSST will, with its unprecedented power for discovering moving 
objects, make a giant leap forward in the studies of the dynamical and chemical 
history of the whole Solar system.

{}

\end{document}